\documentclass[twocolumn]{jpsj2} %% two-column layout
\def\ibid{{\it ibid}.\ }

\def\etal{{\it et\ al.}}

\newcommand{\lsim}
 {\ \raise.35ex\hbox{$<$}\kern-0.75em\lower.5ex\hbox{$\sim$}\ }
\newcommand{\gsim}
 {\ \raise.35ex\hbox{$>$}\kern-0.75em\lower.5ex\hbox{$\sim$}\ }
\def\journal #1#2#3#4{#1 {\bf #2} (#4) #3}

\def\PRP{Phys.\ Rep.}
\def\PRB{Phys.\ Rev.\ B}
\def\PRL{Phys.\ Rev.\ Lett.}

\def\APNY{Ann.\ Phys.\ (New York)}

\def\JPCM{J.\ Phys.\ Cond.\ Mat.}

\def\JPSJ{J.\ Phys.\ Soc.\ Jpn.}

\title{
Nonmonotonic $d_{x^2-y^2}$-Wave Superconductivity in 
Electron-Doped Cuprates Viewing from the Strong-Coupling Side
}% Force line breaks with \\

\author{Tsutomu \textsc{Watanabe}$^{1,2}$\thanks{E-mail: 
h042203d@mbox.nagoya-u.ac.jp}, Takafumi \textsc{Miyata}$^{1}$, 
Hisatoshi \textsc{Yokoyama}$^{3}$, Yukio \textsc{Tanaka}$^{1,2}$ and 
Jun-ichiro \textsc{Inoue}$^{1}$}

\inst{
$^{1}$Department of Applied Physics, Nagoya University,  Nagoya 464-8603 \\
$^{2}$CREST Japan Science and Technology Corporation (JST) \\
$^{3}$Department of Physics, Tohoku University, Sendai 980-8578 \\
}

\abst{
Applying a variational Monte Carlo method to a two-dimensional $t$-$J$ 
model, we study the nonmonotonic $d_{x^2-y^2}$-wave superconductivity, 
observed by Raman scattering and ARPES experiments in the electron-doped 
cuprates. 
As a gap function in the trial state, we extend the $d$-wave form 
(ext.$d$) so as to have its maxima located near the hot spots 
of the system. 
It is found that, in contrast to the hole-doped case, the ext.$d$ wave 
is always more stable than the simple $d$ wave in the electron-doped case, 
and the magnetic correlation of the wave vector ($\pi,\pi$) as well as 
the pair correlation is enhanced. 
These results corroborate spin-correlation-mediated superconductivity 
in cuprates, recently argued from a FLEX calculation. 
In addition, we confirm that $s$- and $p$-wave symmetries are never stabilized 
even in the over-doped regime. 
}

\kword{superconductivity, electron-doped High-$T_{\rm c}$ cuprate, 
nonmonotonic $d_{x^2-y^2}$ wave, antiferromagnetic correlation, hot spot, 
$t$-$J$ model, variational Monte Carlo method}

\begin{document}
\maketitle

%%%%%%%%%%%%%%%%%%%%%%%%%%%%%%%%%%%%%%%%%%%%%%%%%%%%%%%%%%%%%%%%%%%%%%%%%
%\section{\label{sec:intro} Introduction}
%%%%%%%%%%%%%%%%%%%%%%%%%%%%%%%%%%%%%%%%%%%%%%%%%%%%%%%%%%%%%%%%%%%%%%%%%

{\it Introduction}: Antiferromagnetic (AF) correlation is probably 
the primary origin to form Cooper pairs in the high-$T_{\rm c}$ 
cuprates.\cite{Anderson,Scalapino} 
A recent neutron scattering experiment\cite{Yamada} in the electron-doped 
($n$-type) cuprates discovered peaks at magnetic Bragg spots in both normal 
and superconducting (SC) phases, which fact indicates that the spin 
correlation of the AF wave vector ${\bf Q}=(\pi,\pi)$ plays 
an important role in $n$-type cuprates. 
\par

As in the hole-doped ($p$-type) cuprates, the pairing symmetry of the 
$n$-type ones was ascertained to be a $d_{x^2-y^2}$-wave type, as far as 
the doping rate $\delta$ is smaller than 0.15, by a scanning SQUID 
microscope\cite{Tsuei} and angle-resolved photoelectron spectroscopy 
(ARPES).\cite{Sato,Armitage1} 
However, recent experiments by Raman scattering\cite{Blumberg} and 
ARPES\cite{Matsui2} have concluded that the pairing symmetry in the 
$n$-type cuprates is not the typical $d_{x^2-y^2}$ wave characterized by 
$\Delta_{\bf k}\propto\cos k_x-\cos k_y$, but exhibits a non-monotonic behavior. 
Namely, the maximum of $\Delta_{\bf k}$ is located midway between the 
Brillouin-zone boundary $(\pi,0)$ and the zone diagonal $(\pi/2,\pi/2)$. 
Furthermore, for $T>T_{\rm c}$, pseudogap-like behavior\cite{Armitage2} 
or an AF gap\cite{Matsui1} arises at this midway locus, in contrast to 
the $p$-type cuprates, in which the gap maximum and pseudogap behavior 
take place around $(\pi,0)$. 
\par

\begin{figure}
\begin{center}
\includegraphics[width=8.8cm,height=3.8cm]{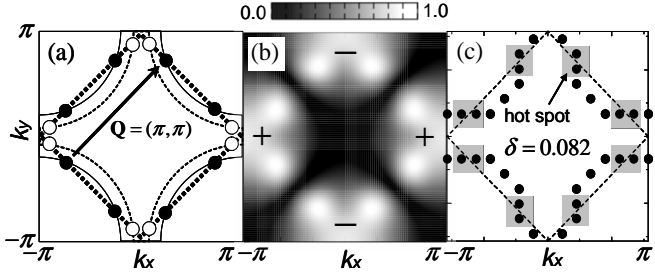}
\end{center}
\vskip -2mm
\caption{
{\bf (a)} Hot spots are compared between a hole-doped ($t'/t=-0.1$, 
$t''/t=0.1$, 
open circle) and an electron-doped ($t'/t=-0.16$, $t''/t=0.2$, solid 
circle) cases for the optimal doping ($\delta=0.15$). 
The bare Fermi surface for the former (latter) parameter set is denoted 
by a thin-dashed (solid) line, and the magnetic Brillouin zone boundary 
by a bold dotted line. 
The AF nesting vector {\bf Q} links two hot spots. 
{\bf (b)} Pairing potential $\Delta_{\bf k}/t$ of the ext.$d$ wave for 
the optimized parameter values (see text) of the system shown in (c). 
{\bf (c)} The $k$-points on the Fermi surface for an underdoped density 
on a $14\times 14$ lattice are shown with solid circles. 
Regions near the hot spots are indicated by shadows, and the magnetic 
Brillouin zone boundary by a dashed line. 
}
\label{fig:hotspot} 
\end{figure}

These results can be explained by the fact that the loci of the gap 
maximum in the p and n types roughly coincide with their respective 
hot spots---the intersection of the Fermi surface and the magnetic 
Brillouin zone boundary [See Fig.~\ref{fig:hotspot}(a)]. 
This observation affords confirmatory evidence that the superconductivity 
in cuprates is induced by the spin correlation (or fluctuation) 
of the wave vector {\bf Q}, which connects the hot spots. 
\par

Theoretically, Yoshimura and Hirashima\cite{Hirashima} recently treated 
this issue, applying a fluctuation exchange approximation (FLEX) 
to the Hubbard model. 
They obtained a non-monotonic behavior of $\Delta_{\bf k}$ and consistent 
results of Raman spectral functions and spin susceptibility etc. with 
the experiments. 
Since the electron correlation is not weak even in the $n$-type cuprates, 
the results of FLEX, which is basically a weak-coupling theory, should 
be checked by complementary studies from the strong-coupling and 
low-carrier-density sides. 
In this paper, we study this issue, applying a variational Monte Carlo 
(VMC) method\cite{VMC} to a $t$-$J$-type model. 
So far, taking this approach, various aspects of the cuprates have been 
elucidated on a strong-coupling footing.\cite{AndersonNew} 
\par

A secondary interest of this paper is possible variation of the 
pairing symmetry from $d$ to $s$ wave in the over-doped regime 
of $n$-type cuprates, observed by tunneling spectroscopy\cite{Biswas} 
and measurement of magnetic penetration depth.\cite{Skinta} 
Although a BCS-level calculation\cite{Khodel} supports these 
experiments, one has to confirm this issue using a less biased 
approach. 
\par

{\it Formulation}: We consider a two-dimensional $t$-$J$ model, 
${\cal H}={\cal H}_t+{\cal H}_J$ with 
$
{\cal H}_t=-\sum_{(i,j)\sigma}t_{ij}{\cal P}_{\rm G}
(c_{i\sigma}^\dagger c_{j\sigma}+{\rm H.c.}){\cal P}_{\rm G} 
$
and 
$
{\cal H}_J=J\sum_{(i,j)}({\bf S}_i\cdot{\bf S}_j
                    -n_in_j/4),
$
%\begin{eqnarray}
% {\cal H}=&-&\sum_{(i,j)\sigma}t_{ij}{\cal P}_{\rm G}
% \left(c_{i\sigma}^\dagger c_{j\sigma}+{\rm H.c.}\right)
%                     {\cal P}_{\rm G}\nonumber \\
%&+&J\sum_{(i,j)}\left({\bf S}_i\cdot{\bf S}_j
%                    -\frac{1}{4}n_in_j\right),
%\end{eqnarray}
where ${\cal P}_{\rm G}=\prod_j(1-n_{j\uparrow}n_{j\downarrow})$, 
the value of $J/t$ is fixed at 0.3, 
and $t_{ij}=t,t',t''$ or 0, according as the site $i$ is a first-, 
second-, third-nearest neighbor or farther site of the site $j$, 
respectively. 
An electron-doped (more-than-half-filled) system can be treated 
with a $t$-$J$ model as a less-than-half-filled case, by applying 
a particle-hole transformation
$c_{j\sigma}^\dagger\rightarrow\exp(i{\bf Q}\cdot{\bf r}_j)h_{j\sigma}$ 
with $t'(t'') \rightarrow -t'(-t'')$. 
Thus, we put $t'/t>0$ $(<0)$ and $t''/t<0$ $(>0)$ for n- (p-)type 
cuprates. 
Actually, we adopt typical values\cite{param} of $t'/t$ and 
$t''/t$ given in the caption of Fig.~\ref{fig:hotspot}. 
\par

%***************************************************************************
%  Fig.1
%***************************************************************************
%******************************************************************************

To this model, we apply a VMC method, which accurately treats 
the local correlation ${\cal P}_{\rm G}$. 
As a variational function for a SC state, a simple Gutzwiller-type 
wave function is used, 
\begin{equation}
|\Psi_{\rm SC}\rangle={\cal P}_{\rm G}|\Phi_{\rm BCS}\rangle=
{\cal P}_{\rm G}\left(\sum_{\bf k}\varphi_{\bf k}
c_{{\bf k}\uparrow}^\dagger c_{{\bf -k}\downarrow}^\dagger
\right)^\frac{N_{\rm e}}{2}|0\rangle, 
\end{equation} 
with 
\begin{equation}
\varphi_{\bf k}=\frac{u_{\bf k}}{v_{\bf k}}=
\frac{\Delta_{\bf k}}
{\varepsilon_{\bf k}-\mu+\sqrt{(\varepsilon_{\bf k}-\mu)^2+\Delta_{\bf k}^2}}.
\end{equation}
Here, $\Phi_{\rm BCS}$ is the BCS wave function of a fixed particle 
number $N_{\rm e}$, 
$\varepsilon_{\bf k}=-2t(\cos k_x+\cos k_y)-4t'\cos k_x\cos k_y
-2t''(\cos 2k_x+\cos 2k_y)$, 
and the parameter $\mu$ is substituted by the chemical potential of 
the non-interacting case. 
It is already known that this type of wave function works well 
for $t$-$J$-type models.\cite{YO,Randeria} 
Anisotropy of the pairing potential is introduced into $\Delta_{\bf k}$, 
as $\Delta_{\bf k}^d=\Delta_0(\cos k_x-\cos k_y)$ for the simple 
$d$ wave. 
We extend it so that it may have large amplitude near the hot spots, 
\begin{eqnarray}
 && \Delta_{\bf k}^{{\rm ext.}d}=\Delta_{\bf k}^d \nonumber \\
 &+&a\Delta_0\Big[k_y^2e^{-c(k_x-b)^2-dk_y^2}  
                   +k_y^2e^{-c(k_x+b)^2-dk_y^2}  \nonumber \\
 &-&k_x^2 e^{-c(k_y-b)^2-dk_x^2}
   -k_x^2e^{-c(k_y+b)^2-dk_x^2}\Big],
\end{eqnarray} 
and call it the extended-$d$ (ext.$d$) wave. 
In $\Delta_{\bf k}^{{\rm ext.}d}$, we add to $\Delta_{\bf k}^d$ 
eight Gaussian peaks, 
whose height, position and width in two directions are adjusted 
by parameter $a$,$b$,$c$ and $d$, respectively. 
Although $\Delta_{\bf k}$ should be determined variationally, 
for simplicity we fix the parameters for the Gaussian, except for 
trial calculations,\cite{note1} at rough optimal values of the 
case $\delta=0.082$ ($L=14$), 
namely, $a=3$, $b=2$, $c=1$ and $d=1$; $\Delta_{\bf k}$ thereof 
is shown in Fig.\ref{fig:hotspot}(b). 
Note that the loci of the maximal $\Delta_{\bf k}$ are situated 
very closely to the hot spots, shown in Fig.\ref{fig:hotspot}(c). 
Thus, $\Delta_0$ (amplitude of $\Delta_{\bf k}$) becomes a sole 
parameter to be optimized. 
Unlike the BCS theory, $\Delta_{\bf k}$ here is not entirely 
equivalent to the SC gap, especially in $\Delta_0$, but the 
symmetry of $\Delta_{\bf k}$ faithfully reflects the SC gap. 
\par

We have followed a conventional VMC scheme,\cite{VMC} and collected 
$10^5$-$10^6$ samples, which suppress the statistical errors in energy 
at $\sim 10^{-4}t$. 
The system used has $L\times L$ sites ($L=12, 14$ and 16) with the 
periodic-antiperiodic boundary conditions. 
The particle density $\delta$ is chosen so as to satisfy the 
closed-shell condition. 
\par

\begin{figure}
\begin{center}
\includegraphics[width=8.7cm,height=5.8cm]{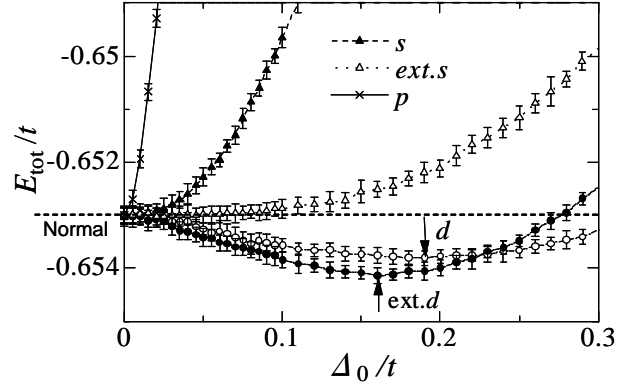}
\end{center}
\vskip -6mm
\caption{
Total energies versus a variational parameter $\Delta_0$ for various 
pairing symmetries for an electron-doped system with $\delta =0.163$. 
The arrows on the $d$ and ext.$d$ waves denote the optimal values. 
The value of the normal state ($\Delta_0=0$) is shown with a dashed line. 
The system is of $14 \times 14$, and $J/|t|=0.3$.
}
\label{fig:spetot} 
\end{figure}

{\it Results}: Before going to the $d$-type pairings, we check 
the stability of the singlet $s$- ($\Delta_{\bf k}=\Delta_0$), 
ext.~$s$-[$=\Delta_0(\cos k_x+\cos k_y)$] and triplet $p$-wave 
($=\Delta_0 \sin k_x$) symmetries. 
In Fig.~\ref{fig:spetot}, the variational energies $E_{\rm tot}$ 
of the symmetries we treat are compared in the overdoped regime. 
Here, $E_{\rm tot}$'s of the $s$, ext.$s$ and $p$ waves monotonically 
increase, as $\Delta_0$ increases, and are destabilized with respect 
to the normal state. 
We have also confirmed that the same behavior persists to a large 
doping rate, $\delta=0.245$, where SC states are no longer stabilized. 
Since our treatment is little biased in comparing pairing symmetries, 
we are confident that the $p$ and $s$-type waves are not realized 
even in the overdoped regime; thus a pairing-symmetry transition 
is unlikely to arise.\cite{note2} 
\par

%***************************************************************************
%  Fig.2
%***************************************************************************
%******************************************************************************

\begin{figure}
\begin{center}
\includegraphics[width=8cm,height=9cm]{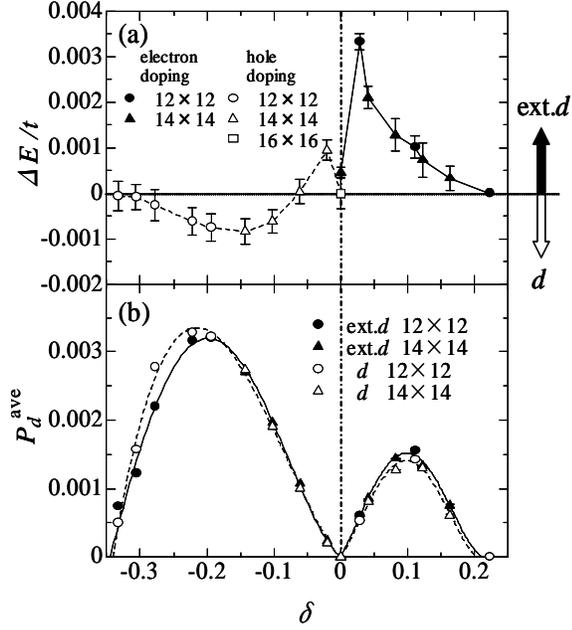}
\end{center}
\vskip -6mm
\caption{
{\bf (a)} Difference in the optimized energy between the ext.$d$- and 
simple $d$-wave states, 
$\Delta E = E^d_{\rm tot}-E^{{\rm ext.}{d}}_{\rm tot}$, 
as a function of doping rate $\delta$. 
Since the parameters ($t'$ and $t''$) are different between p and 
n types, the values at $\delta=0$ do not coincide. 
{\bf (b)} Comparison of the pair correlation function between the ext.$d$ 
and simple $d$ waves as a function of $\delta$. 
In both panels, hole-doped systems are indicated by negative values of 
$\delta$. 
}
\label{fig:differ} 
\end{figure}

Now, we turn to the $d$-type symmetries. 
By contrast, the variational energies of the $d$- and ext.$d$-wave SC 
states plotted in Fig.~\ref{fig:spetot} have minima at finite values 
of $\Delta_0$, as indicated by arrows. 
The decrease in $E_{\rm tot}$ of the $d_{x^2-y^2}$ wave has been 
well-known for the plain $t$-$J$ model ($t'=t''=0$) since the early 
stage.\cite{earlyt-J}
It should be noted here that the ext.$d$ wave has an appreciably lower 
energy than the simple $d$ wave even for such a large value of $\delta$. 
We have optimized $E_{\rm tot}$ similarly for various values of $\delta$, 
and depict in Fig.\ref{fig:differ}(a) the difference of $E_{\rm tot}$ 
between the $d$ and ext.$d$ waves, 
$\Delta E = E^d_{\rm tot}-E^{{\rm ext.}{d}}_{\rm tot}$.\cite{notedelta} 
In electron-doped cases, the ext.$d$ wave is always more stable than 
the $d$ wave. 
The large value of $\Delta E$ near half filling probably stems 
from the fact that the Gaussian peaks in $\Delta_{\bf k}^{{\rm ext.}d}$ 
are close to the hot spots, as well as that the energy scale in the 
condensation energy $E^{\rm cond} 
(=E^{\rm normal}_{\rm tot}-E^{\rm SC}_{\rm tot}$) 
becomes large for $\delta\rightarrow 0$. 
$\Delta E$ vanishes at $\delta=0.222$, where the hot spots still survive, 
but $E^{\rm cond}$ vanishes for both waves. 
Conversely, in the hole-doped cases [open symbols in Fig.\ref{fig:differ}(a)], 
the simple $d$ wave is more stable than the ext.$d$ wave except in the 
vicinity of half filling. 
In the optimal- and overdoped regime for $\delta<0$, the hot spots sit 
near ($\pi,0$) and equivalent points, whereas near half filling 
the hot spots are still away from ($\pi,0$) point and rather closer to 
the Gaussian positions we set [Fig.\ref{fig:hotspot}(b)]. 
Thus, the results in $E_{\rm tot}$ definitely indicate that the SC state 
becomes stable when the maximum of the gap $\Delta_{\bf k}$ is located 
near the hot spots of the system. 
\par

%***************************************************************************
%  Fig.3
%***************************************************************************

%***************************************************************************

To reinforce the above argument, we consider the $d$-wave SC correlation 
function of nearest-neighbor pairs, 
$$
P_d({\bf r})=\frac{1}{N_{\rm s}}
\sum_{\bf j}\sum_{\tau,\tau'={\bf x},{\bf y}}
(-1)^{1-\delta(\tau,\tau')}
\langle\Delta_\tau({\bf j})\Delta_{\tau'}({\bf j}+{\bf r})\rangle,
$$
with 
$
\Delta^\dagger_\tau({\bf j})=
(c^\dagger_{{\bf j}\uparrow}c^\dagger_{{\bf j}+\tau\downarrow}+
 c^\dagger_{{\bf j}+\tau\uparrow}c^\dagger_{{\bf j}\downarrow}).
$
Since $P_d({\bf r})$ rapidly decays with $|{\bf r}|$ and is almost 
constant for $|{\bf r}|\ge 3$, the average for $|{\bf r}|\ge 3$, 
$P_d^{\rm ave}$, gives an adequate estimate of the long-distance value. 
In Fig.\ref{fig:differ}(b), $P_d^{\rm ave}$ is plotted versus 
carrier density. 
The remarkable asymmetry between the p and n types can be attributed 
to the difference in DOS at the Fermi surface,\cite{Shih} particularly 
at the hot spots. 
Note that the ext.$d$ wave always exceeds the $d$ wave for the n type, 
whereas the relation is inverse in the over-doped regime for the 
p type. 
This tendency of $P_d({\bf r})$ corresponds well with that of 
$\Delta E$ [Fig.\ref{fig:differ}(a)]. 
Incidentally, in $n$-type systems, the SC correlation of 
long-distance pairs is probably enhanced, as will be discussed later.
\par

%***************************************************************************
%  Table I
%***************************************************************************
\begin{table}
\caption{
Comparison of the energy components, $E_t$ and $E_J$, for four carrier 
densities with $J/t=0.3$ and $t$ being the unit.
The energy components of the simple $d$ wave, $E_t^d$ and $E_J^d$, are 
entered in the first two lines. 
Entered in the middle and lower two lines are the differences of $E_t$ 
and $E_J$ between the $d$-wave and the normal state and between the $d$ 
and ext.$d$ waves, respectively. 
}
\label{table:1}
\begin{tabular}{ccccc} \hline
\quad$\delta$  & $-0.143$ & $-0.061$ & 0.082 & 0.163 \\ \hline
$E_t^d$ & $-0.3541(1)$ & $-0.1634(1)$ & $-0.2314(1)$ & $-0.4593(1)$ \\
$E_J^d$ & $-0.2336(1)$ & $-0.2876(0)$ & $-0.2566(1)$ & $-0.1946(0)$ \\ \hline
$\Delta E_t^{\rm N}$ & 0.0090(5) & 0.0066(4) & 0.0151(2) & 0.0043(2) \\
$\Delta E_J^{\rm N}$ & $-0.0288(2)$ & $-0.0471(2)$ & $-0.0313(2)$ & $-0.0052(1)$ \\ \hline
%$\Delta_0^{{\rm ext.}d}/t$ & 0.28 & 0.47 & 0.47 & 0.16 \\
$\Delta E_t$ & $-0.0005(2)$ & $-0.0008(2)$ & $-0.0008(2)$ & $-0.0014(2)$ \\
$\Delta E_J$ & $-0.0004(1)$ &  0.0008(1) &  0.0021(1) &  0.0018(1) \\ \hline
\end{tabular}

\end{table}
%***************************************************************************

Next, to identify the origin of the energy gain in $E_{\rm tot}$, 
we compare the energy components, namely hopping energy 
$E_t=\langle{\cal H}_t\rangle$ and exchange energy 
$E_J=\langle{\cal H}_J\rangle$, 
among the $d$-wave, ext.$d$-wave and normal states. 
In Table \ref{table:1}, we list the raw values of the $d$-wave, 
$E_t^d$ and $E_J^d$, and the differences between them and those of 
the other two states, namely, $\Delta E_t^N(=E_t^d-E_t^{\rm normal})$, 
$\Delta E_t(=E_t^d-E_t^{{\rm ext.}d})$, etc.~for four kinds of doping. 
As compared with the normal state, the $d$- (also ext.$d$-) wave SC 
state is stabilized by the noticeable decrease in $E_J$; conversely, 
$E_t$ is more or less increases. 
Such magnetic origin of superconductivity is characteristic of 
$t$-$J$-type models,\cite{YO,magnetic} and closely related to the 
kinetic-energy-driven SC mechanism in the strong-correlation regime 
of the Hubbard model.\cite{YTOT} 
In comparison of the $d$ and ext.$d$ waves, $E_J^{{\rm ext.}d}$ is 
lower whenever the ext.$d$ wave has a lower total energy 
($\delta\gsim-0.06$), whereas the ext.$d$ wave always possesses 
somewhat higher $E_t$. 
Thus, the gap structure of the ext.$d$ wave has further advantages 
to gain magnetic energy for $n$-type cases over the simple $d$ wave. 
\par 

%***************************************************************************
%  Fig.4
%***************************************************************************
\begin{figure}
\begin{center}
\includegraphics[width=8.5cm,height=4cm]{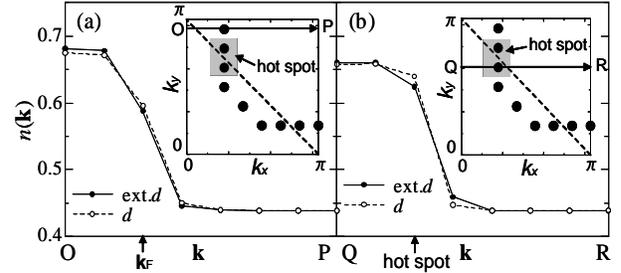}
\end{center}
\vskip -3mm
\caption{
Comparison of the momentum distribution function between the ext.$d$ and 
pure $d$ waves along the path indicated by an arrow in each inset, 
for the electron-doped system with $\delta =0.122$. 
In each inset, solid dots and a broken line denote the outmost occupied 
$k$-points (the Fermi surface) due to $\varepsilon_{\bf k}$ and the
AF Brillouin zone boundary, respectively. 
The system is of $14 \times 14$. 
}
\label{fig:nkhot2} 
\end{figure}
%***************************************************************************

Now, we consider the momentum distribution function, 
$n({\bf k})=1/2\sum_\sigma
\langle c_{{\bf k}\sigma}^\dagger c_{{\bf k}\sigma}\rangle$, 
to actually observe the gap behavior in the momentum space---a milder 
slope at quasi-$k_{\rm F}$.
In Figs.~\ref{fig:nkhot2}(a) and (b), we depict $n({\bf k})$ for 
an electron-doped case\cite{notenk} along two paths, namely 
OP [in (a)], which goes away from the hot spots, and QR [in (b)], 
which penetrates the hot-spot area. 
In (a), $n({\bf k})$'s for the $d$ and ext.$d$ waves exhibit 
similar behavior, and the $d$ wave seems slightly mild. 
On the other hand, in (b) the ext.$d$ wave makes an obviously milder 
curve around the hot spot. 
Thus, the gap behavior around the hot spot is enhanced
in the ext.$d$-wave state. 
Incidentally, in the node-of-gap direction, (0,0)-($\pi,\pi$), 
the difference between the $d$ and ext.$d$ waves is very small, 
and a clear Fermi surface [discontinuity in $n({\bf k})$] can be 
seen (not shown). 

%***************************************************************************
%  Fig.5
%***************************************************************************
\begin{figure}
\begin{center}
\includegraphics[width=8.7cm,height=4.7cm]{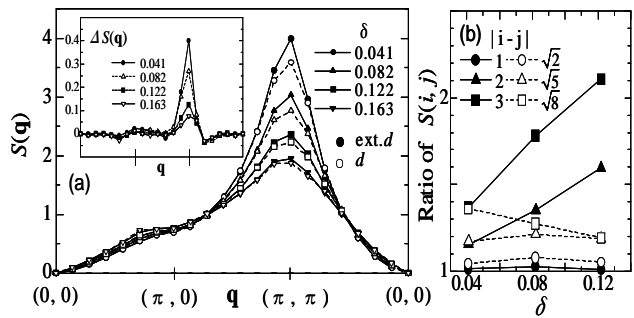}
\end{center}
\vskip -5mm
\caption{
{\bf (a)} Comparison of the spin structure factor between the ext.$d$ 
and pure $d$ waves along the path 
$(0,0)\rightarrow(\pi,0)\rightarrow(\pi,\pi)\rightarrow(0,0)$  
for four densities of electron doping. 
Plotted in the inset is the difference of $S({\bf q})$ between the 
ext.$d$ and pure $d_{x^2-y^2}$ waves, 
$\Delta S({\bf q})=S^{{\rm ext.}{d}}({\bf q})-S^d({\bf q})$. 
{\bf (b)} Ratio of real-space spin correlation function between 
the $d$ and ext.$d$ waves, 
$S^{{\rm ext.}d}({\bf i},{\bf j})/S^d({\bf i},{\bf j})$, is plotted 
for several values of $|{\bf i-j}|$ as a function of $\delta$. 
The data shown are the average among the four (or eight) equidistant 
sites. 
The system is of $14 \times 14$. 
}
\label{fig:sq_sij} 
\end{figure}
%******************************************************************************

Finally, we consider the spin correlation function. 
In Fig.\ref{fig:sq_sij}(a), the spin structure factor, 
$S({\bf q})=1/N_{\rm S}\sum_{\bf ij}{e^{i{\bf q}\cdot({\bf i-j})}
S({\bf i},{\bf j})}$ with 
$S({\bf i},{\bf j})=\langle S_{\bf i}^z S_{\bf j}^z\rangle$, 
of the $d$ and ext.$d$ waves is plotted for several densities of 
electron doping. 
Both waves have the maximal amplitude at ($\pi,\pi$) for all the 
electron densities, which is consistent with the neutron 
experiment.\cite{Yamada}
As shown in the inset of Fig.\ref{fig:sq_sij}(a), the difference 
between the $d$ and ext.$d$ waves is almost restricted to the 
vicinity of ($\pi,\pi$), where $S(\bf q)$ of the ext.$d$ wave 
is sizably enhanced. 
This enhanced AF correlation naturally leads to the energy gain 
in $E_J$, mentioned above. 
Shown in Fig.\ref{fig:sq_sij}(b) is the ratio of the real-space spin 
correlation function for the ext.$d$ wave to that for the simple 
$d$ wave. 
Although the $d$ and ext.$d$ waves exhibit almost the same 
values for the nearest-neighbor sites ($|{\bf i}-{\bf j}|=1$), 
$S({\bf i},{\bf j})$ of the ext.$d$ wave for farther distances 
considerably increases, especially in the $x$ (or $y$) direction. 
This is because the harmonics of the $d$ wave, 
$\cos nk_x-\cos nk_y$ ($n\ge 2$), give substantial contribution 
to $\Delta_{\bf k}^{{\rm ext.}d}$.\cite{Hirashima} 
Unlike the pure $d$ wave ($n=1$), they elongate the coherence 
length (and magnetic correlation length);  
thereby the correlation strength in the electron-doped systems 
becomes effectively weaker.\cite{Yanase} 
Such tendency has been actually observed by various 
experiments.\cite{weak} 
\par

{\it Summary}: Using a variational Monte Carlo method 
for a $t$-$J$ model, we have studied a nonmonotonic (ext.) $d$-wave 
superconducting state, in which the amplitude of the gap parameter 
$\Delta_{\bf k}$ is intentionally enhanced around the hot spots,  
so as to agree with recent experiments of Raman scattering and ARPES. 
This ext.$d$-wave state has an appreciably lower energy than the 
simple $d$-wave state for all the densities of electron doping (also 
very low hole doping). 
This stabilization of the ext.$d$ wave is caused by the gain in 
magnetic exchange energy, accompanied by a marked increase in the 
spin correlation of the wave vector ${\bf Q}=(\pi,\pi)$. 
In addition, we have shown that the $s$-type and $p$ waves are 
unlikely to take place in the high-$T_{\rm c}$ regime of $\delta$.
Our results using a strong-coupling approach basically agree 
with the recent FLEX study;\cite{Hirashima} 
thereby, it is ensured that the AF spin correlation 
plays a crucial role for the high-$T_{\rm c}$ superconductivity. 
\par

We believe that the essence of the nonmonotonic $d$-wave gap is 
grasped in this work, although we have simplified 
$\Delta_{\bf k}^{{\rm ext.}d}$ by both a naive assumption of 
a Gaussian form and fixing the parameters controlling the Gaussian.
We should address quantitative refinement as well as related 
issues. 
For instance, (1) simultaneous optimization of all the variational 
parameters, including the renormalization of the quasi-Fermi surface, 
enables us to follow the continuous evolution of $\Delta_{\bf k}$ 
versus $\delta$. 
(2) The relation to the AF ordered state is very important.

%-----------------------------------------------------------------------

{\it Acknowledgments}:
One of the authors (H.Y.) appreciates useful discussions with M.~Ogata.
This work is partly supported by Grant-in-Aids from the Ministry of 
Education etc., by the Supercomputer Center, ISSP, University of Tokyo, 
by the 21st Century COE "Frontiers of Computational Science", 
and by NAREGI Nanoscience Project. 
\par

%%%%%%%%%%%%%%%%%%%%%%%%%%%%%%%%%%%%%%%%%%%%%%%%%%%%%%%%%%%%%%%%%%%%%%%%%

%%%%%%%%%%%%%%%%%%%%%%%%%%%%%%%%%%%%%%%%%%%%%%%%%%%%%%%%%%%%%%%%%%%%%%%%%


\begin{thebibliography}{99}
\bibitem{note1} The variational energy weakly depends on $a$,$c$ and 
$d$, but has a relatively narrow minimum as a function of $b$. 
Thus, the position of the Gaussian peaks has a special significance.
\bibitem{note2} 
Experiments on the surface have yielded both affirmative \cite{Chesca} 
and contradictory\cite{Biswas,Skinta,s-wave} conclusions to 
the $d$-type waves. 
They may be affected by other factors than electronic origin, 
for instance, surface impurities.\cite{Asano}
\bibitem{notedelta} We use a negative value of $\delta$ for a 
hole-doped case so as to discuss the p and n types of doping 
simultaneously.
\end{thebibliography}
\end{document}